\documentstyle[aps,prb,epsf,floats]{revtex}

\draft

\begin{document}
\twocolumn[\hsize\textwidth\columnwidth\hsize\csname@twocolumnfalse%
\endcsname

\title {A fractional-spin phase in the power-law Kondo model}

\author{Matthias Vojta and Ralf Bulla}
\address{Theoretische Physik III, Elektronische Korrelationen und
Magnetismus, Institut f\"ur Physik \\
Universit\"at Augsburg, 86135 Augsburg, Germany}
\date{Feb 17, 2002}
\maketitle

\begin{abstract}

We consider a Kondo impurity coupled to a fermionic host
with a power-law density of states near the Fermi level,
$\rho(\epsilon) \sim |\epsilon|^r$, with exponent $r<0$.
Using both perturbative renormalization group (poor man's scaling)
and numerical renormalization group methods,
we analyze the phase diagram of this model
for ferromagnetic and antiferromagnetic Kondo coupling.
Both sectors display non-trivial behavior with several
stable phases separated by continuous transitions.
In particular, on the ferromagnetic side there is a
stable intermediate-coupling fixed point with universal properties
corresponding to a fractional ground-state spin.

\end{abstract}
]

\section{Introduction}

The low-temperature physics of the standard Kondo model,
describing a single magnetic impurity embedded in a metal,
is by now well understood~\cite{hewson}.
For antiferromagnetic coupling between the impurity and
the conduction electron spins
the effective interaction grows with decreasing temperature.
The low-energy behavior 
is completely determined by a single energy scale,
the Kondo temperature $T_{\rm K}$,
and the impurity spin is fully quenched in the low-temperature limit,
$T \ll T_{\rm K}$.
In contrast, for ferromagnetic coupling the
effective interaction decreases upon lowering the temperature,
leaving the impurity spin essentially decoupled from the environment.

The standard Kondo picture has to be revised
if the conduction band density of states (DOS)
is not constant near the Fermi level.
This is the case in systems with a
power-law DOS $\rho(\epsilon) \sim |\epsilon|^r$.
Exponents $r>0$ lead to a vanishing DOS
at the Fermi level --
such a pseudogap DOS arises in one-dimensional interacting systems,
in certain zero-gap semiconductors,
and in systems with long-range order where
the order parameter has nodes at the Fermi surface, {\em e.g.},
$p$- and $d$-wave superconductors ($r=2$ and 1).
The pseudogap Kondo problem has attracted a lot of attention
during recent
years~\cite{withoff,largeN,cassa,chen,bulla,ingersent,GBI,bulla2,pgko}.
These studies show the existence of a zero-temperature boundary
phase transition at a critical antiferromagnetic Kondo
coupling, $J_c$, below which the impurity spin is unscreened
even at lowest temperatures.
A comprehensive discussion of possible fixed points
and their thermodynamic properties has been given by Gonzalez-Buxton and
Ingersent \cite{GBI} based on the numerical renormalization group (NRG) approach.

In this work we investigate a power-law Kondo model
for exponents $-1<r<0$ which has not been explored before.
Negative $r$ implies a diverging low-energy DOS,
corresponding to a critical or van-Hove singularity at the
Fermi level.

The Kondo Hamiltonian for a spin-1/2 impurity can be written as
$H=H_{\rm band}+H_{\rm int}$, with
\begin{equation}
H_{\rm int} = J {\bf S} \cdot {\bf s}_0 + V c_{0\sigma}^\dagger c_{0\sigma}
\end{equation}
and
$H_{\rm band} = \sum_{{\bf k}\alpha} \epsilon_{\bf k} c^\dagger_{{\bf k}\alpha} c_{{\bf k}\alpha}$
in standard notation,
${\bf s}_0 = \sum_{\bf kk'\alpha\beta} c^\dagger_{{\bf k}\alpha} {\bf \sigma}_{\alpha\beta}
c_{{\bf k '}\beta}$ is the conduction band spin operator at the impurity
site ${\bf r}_0 = 0$, and ${\bf S}$ denotes the impurity spin operator.
For simplicity we will use a conduction band with
a symmetric density of states,
$\rho(\epsilon) = \rho_0 |\epsilon/D|^r$ for $|\epsilon| < D = 1$,
$\rho_0 = (1+r)/(2D)$.
Particle-hole asymmetry is introduced via the potential scattering
term $V$.

\begin{figure}[!t]
\epsfxsize=3.4in
\centerline{\epsffile{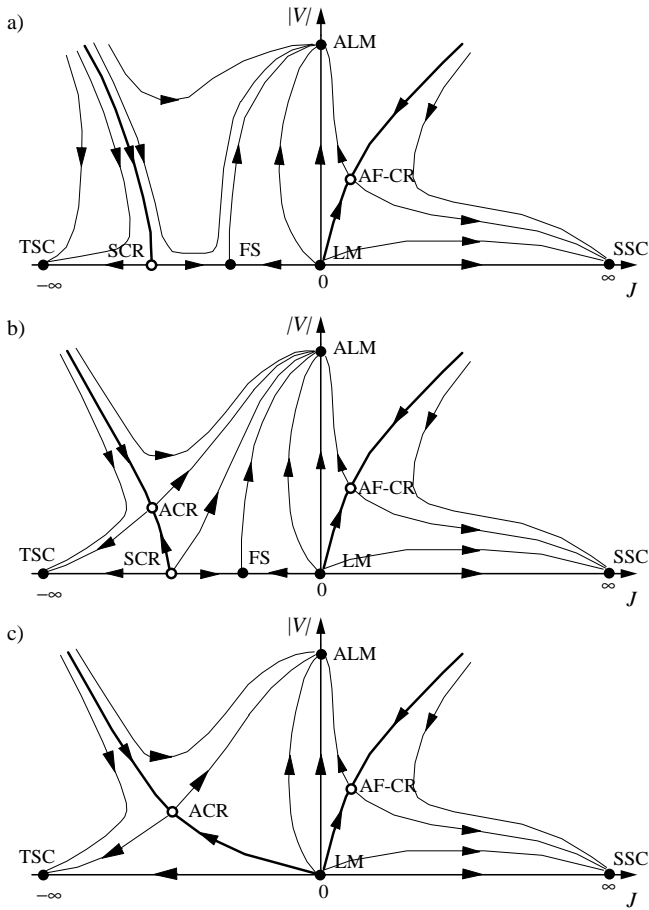}}
\vspace*{5pt}
\caption{
Schematic renormalization group flows for
the power-law Kondo model with $r<0$,
as deduced from NRG calculations (see text).
a) $\bar{r}^\ast<r<0$,
b) $\bar{r}_{\rm max}<r<\bar{r}^\ast$,
c) $-1<r<\bar{r}_{\rm max}$,
with NRG estimates of
$\bar{r}^\ast = -0.245 \pm 0.005$,
$\bar{r}_{\rm max} = -0.265 \pm 0.005$.
The solid dots denote infrared stable fixed points,
the open dots are critical fixed points
(labelled as in the text).
FS denotes the infrared stable intermediate-coupling
fixed point with fractional spin.
}
\label{figflow}
\end{figure}

Our findings (Fig.~\ref{figflow}) can be summarized as follows:
Antiferromagnetic Kondo interactions in the particle-hole symmetric
model ($V\!=\!0$) always flow to a
strong-coupling fixed point with a singlet ground state (SSC fixed point)
and complete screening of the impurity.
The ferromagnetic sector has very rich behavior: large interactions
flow to a {\em ferromagnetic} strong-coupling fixed point with a
triplet ground state (TSC).
Even more interestingly, for $V\!=\!0$
there exists a {\em stable} intermediate-coupling
fixed point (FS).
Its most remarkable property is a Curie response
corresponding to a fractional spin,
\begin{equation}
\chi_{\rm imp} (T\to 0) = \frac{{\cal C}(r)}{T}\,,
\label{curie}
\end{equation}
where $\chi_{\rm imp}$ is the total impurity
contribution to the uniform susceptibility,
and $\cal C$ is a universal, irrational number
which depends only on the DOS exponent $r$.
(The $V\!\neq\!0$ behavior is quite complicated as well and
will be discussed below.)

Such fractional-spin states at intermediate coupling
have so far only been found at infrared unstable
(critical) fixed points, namely in the pseudogap Kondo model
($r>0$) \cite{GBI}
and for impurities coupled to quantum-critical magnets \cite{impmag}.
We note that ``exotic'' states corresponding to stable intermediate-coupling
fixed points also occur in multichannel Kondo models \cite{hewson,NB,bullafp};
here, however,
the leading term in the impurity susceptibility does not follow a Curie law.


\section {Weak-coupling analysis}

We start our considerations with the standard weak-coupling
renormalization group treatment (poor man's scaling \cite{poor}), here
modified for the power-law Kondo model.

Integrating out a shell of high-energy conduction electrons gives
the one-loop renormalization of the Kondo interaction of order $J^2$,
rescaling of energies and band cut-off $\Lambda$ leads to an additional renormalization
of all couplings proportional to $r$~\cite{withoff}.
Expressed in $\beta$ functions for the dimensionless running couplings
$j=\rho_0 J$ and $v=\rho_0 V$ we have
\begin{eqnarray}
\frac{d j}{d \ln \Lambda} = j (r - j) + {\cal O}(j^3) ~,~~~~
\frac{d v}{d \ln \Lambda} = r v .
\label{rgeq}
\end{eqnarray}
For $r=0$ the potential scattering term $v$ remains unchanged
whereas positive (negative) Kondo coupling $j$ flows to $\infty$ (zero)
upon decreasing the cut-off.
In the pseudogap case, $r>0$, small $j$ of either sign renormalizes
to zero.
For small positive $r$ one can deduce a critical fixed point at $J_c = r/\rho_0$, $V_c = 0$,
corresponding to the transition between a local moment phase and
an antiferromagnetic strong coupling phase in the pseudogap
model \cite{withoff}.

\begin{figure}[!t]
\epsfxsize=3.3in
\centerline{\epsffile{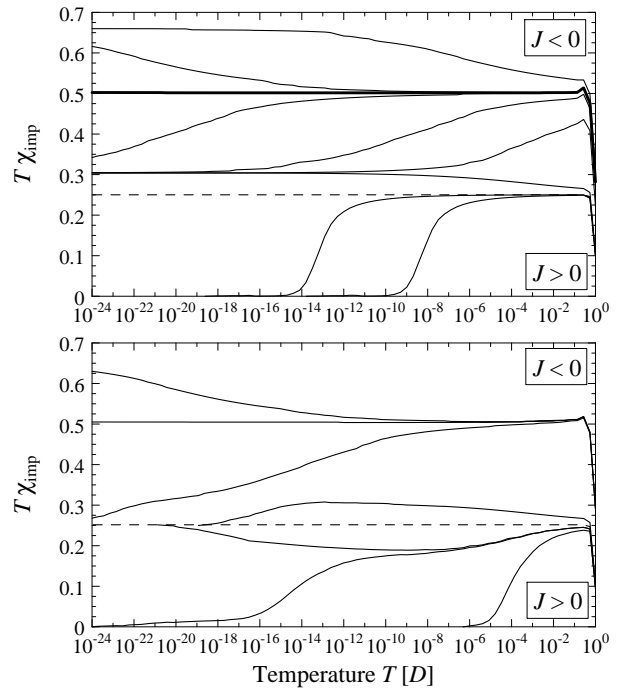}}
\vspace*{5pt}
\caption{
NRG results \protect\cite{nrgprob}
for $T\chi_{\rm imp}$ ($k_B\!=\!1$)
of the power-law Kondo model with $r\!=\!-0.2$,
illustrating the flow in Fig.~\protect\ref{figflow}a.
The dashed curve shows a free spin $J\!=\!0$
for comparison.
a) Particle-hole symmetric case $V\!=\!0$.
The thick line corresponds to the critical point at
$J_c \simeq -5.6302$.
The other curves are for (top to bottom)
$J =$ $-7$, $-5.65$, $-5.6$, $-5$, $-3$, $-0.1$, $+0.001$, $+0.01$.
b) Small asymmetry $V\!\neq\!0$.
Shown are (top to bottom)
$(J,V)=$ $(-7,0.75)$, $(-7,0.75735)$, $(-7,0.8)$, $(-0.1,0.001)$,
$(0.045,0.1)$, $(0.046,0.1)$ $(0.1,0.1)$.
}
\label{fignrg1}
\end{figure}

Now we turn to $r<0$:
here small $j$ grows for {\em both} signs of the Kondo coupling.
In analogy to the above,
on the ferromagnetic side for small negative $r$, we can predict a fixed point
at $J^* = - |r|/\rho_0$ which is now infrared stable (for $V=0$).
The properties of this novel intermediate-coupling fixed point
are perturbatively accessible in a double expansion in $j$ and $r$ --
as known from the theory of critical phenomena the properties are
determined by the {\em universal} value of the running coupling constant $j$,
and are completely different from the properties of usual stable phases where
the renormalized couplings are either zero or infinite.
In particular, local spin correlations show a power law
decay with some non-trivial, $r$-dependent exponent, the
ground-state entropy has a universal, non-zero value,
and the total impurity-induced susceptibility shows
fractional Curie behavior (\ref{curie}).
In principle the calculation of these properties can be done in
renormalized perturbation theory up to high orders;
this does, however, not provide information
about the fixed point structure for {\em finite}
negative $r$ nor about the flow of large initial couplings $J$.

\section {Numerical results}

To verify the above picture and investigate the strong-coupling behavior,
we have performed extensive studies of the power-law Kondo model
using the NRG technique \cite{nrg}.
From the flow of the NRG energy levels
we have deduced a number of (meta)stable
fixed points\cite{GBI}, with properties listed in the following:

\noindent
(LM)  The symmetric local-moment fixed point corresponds
to the system with a decoupled impurity, $J=V=0$.
The $T\!=\!0$ limit of the susceptibility is $T \chi_{\rm imp} = 1/4$,
the entropy $S_{\rm imp} =\ln 2$.
This fixed point is unstable both w.r.t. to
finite $J$ and $V$, which follows from (\ref{rgeq}).
The instability against potential scattering can be
easily understood:
finite $V$ leads to a pole in the scattering T matrix which
creates a local (anti)bound state, leading to the
ALM behavior below.

\noindent
(ALM) An additional asymmetric local-moment fixed
point exists, corresponding to $J\!=\!0$ and $|V|\!=\!\infty$;
here the conduction electron site next to the impurity is
either doubly occupied or empty.
The impurity thermodynamic properties are similar to the
LM fixed point.

\noindent
(SSC) The singlet strong-coupling fixed point is
reached for positive $J$ and small $|V|$,
it corresponds to $J\!=\!\infty$, $V\!=\!0$ and displays a fully
quenched spin with $T \chi_{\rm imp} = 0$,
$S_{\rm imp}=0$.
(The leading term in $\chi_{\rm imp}$ has the form $T^{-r}$, {\em i.e.},
the impurity susceptibility itself vanishes in the zero-temperature
limit.)
In contrast to the power-law model with $r>0$, this fixed point is
stable w.r.t. particle-hole symmetry breaking.

\noindent
(TSC) A triplet strong-coupling fixed point is
reached for large negative $J$,
its properties follow from $J\!=\!-\infty$, $V\!=\!0$ as
$T \chi_{\rm imp} = 2/3$, $S_{\rm imp}=\ln 3$.
This fixed point is also stable against
finite $V$.

\noindent
(FS) The new, fractional-spin fixed point exists
for small $r<0$ and is reached for small $J<0$, $V=0$.
It features universal, $r$-dependent values
of $T \chi_{\rm imp}$ and $S_{\rm imp}$; it is unstable
against non-zero potential scattering.

From the NRG results we can construct the flow diagrams
shown in Fig.~\ref{figflow}, and we can obtain various
thermodynamic quantities as function of $T$ \cite{GBI}.
In Fig.~\ref{fignrg1} we display the temperature dependence of
$T\chi_{\rm imp}$ for different Kondo couplings $J$
and a fixed DOS exponent $r=-0.2$, which puts the model
in the intervall ${\bar r}^\ast < r < 0$, with
the flow diagram depicted in Fig.~\ref{figflow}a.
%
%
The particle-hole symmetric case $V\!=\!0$ is shown in Fig.~\ref{fignrg1}a.
Positive values of $J$ lead to a fully quenched spin in the
low-temperature limit.
The ``Kondo'' temperature characterizing
the flow to strong coupling depends in a power-law fashion
on $J$, $T_K \sim D (\rho_0 J)^\alpha$, with the leading
term being $\alpha = 1/r$.
The behavior is in contrast to
the exponential dependence $T_K(J)$ in the metallic $r=0$ case,
reflecting the fact that for $r<0$ antiferromagnetic $J$ is a relevant
rather than a marginally relevant perturbation of the local-moment fixed
point.

Turning to negative values of $J$, we see that all small
ferromagnetic values of the initial coupling yield a flow
to the predicted intermediate-coupling fixed point (FS),
characterized by a universal
Curie behavior of the impurity susceptibility,
here $T \chi_{\rm imp} \approx 0.30$.
Interestingly, the basin of attraction of this fixed point does
not extend to arbitrarily large $|J|$: large ferromagnetic
couplings flow to a different stable fixed point with
$T \chi_{\rm imp} = 2/3$ -- this is the triplet strong-coupling (TSC) fixed
point.
The boundary quantum phase transition between the fractional-spin
phase and the triplet phase appears to be continuous,
and consequently there is a critical
particle-hole symmetric fixed point (SCR, thick line in Fig.~\ref{fignrg1}a),
separating the fractional-spin and triplet strong-coupling regimes.

Switching on potential scattering, Fig.~\ref{fignrg1}b,
we find that the local-moment and fractional-spin fixed points are unstable
w.r.t. particle-hole symmetry breaking,
and both yield flows towards the asymmetric local moment (ALM) fixed point.
(Note, however, that $T\chi_{\rm imp}=1/4$ for both
the symmetric and asymmetric local-moment fixed points.)
In contrast, both strong-coupling fixed points
as well as the critical fixed point (SCR) are stable
w.r.t. finite $V$ (Fig.~\ref{figflow}a).
On the antiferromagnetic side, finite $V$ suppresses Kondo screening,
leading to another non-trivial transition between the singlet strong-coupling (SSC)
and asymmetric local-moment (ALM) phases, with an associated particle-hole
asymmetric critical fixed point at positive $J$
(AF-CR, here $T \chi_{\rm imp} \approx 0.18$ in Fig.~\ref{fignrg1}b).

Our numerical investigation shows that
the described fixed point structure on the ferromagnetic side changes
for more negative $r$ values.
With decreasing $r<0$, first the symmetric critical fixed point between
the triplet strong-coupling (TSC) and fractional-spin (FS) phases
becomes unstable w.r.t. broken particle-hole symmetry,
and a second critical fixed point (ACR) with finite asymmetry
appears.
This means that at a value $r = {\bar r}^\ast$ the critical fixed
point splits, and the transition between triplet strong-coupling (TSC)
and asymmetric local-moment phases (ALM) is now
governed by the new asymmetric critical point ACR.
The resulting renormalization group flow is shown in
Fig.~\ref{figflow}b.
(Strictly speaking, each described fixed point at
finite $|V|$ represents a pair of fixed points,
which are trivially related by a particle-hole
transformation.)

Upon further decreasing $r$, the ferromagnetic
particle-hole symmetric critical fixed point SCR
moves to smaller couplings, towards the stable FS fixed point.
At a certain value $r = {\bar r}_{\rm max}$ both fixed points meet
and disappear (!).
This implies that for $r < {\bar r}_{\rm max}$ and $V=0$
any negative value of $J$ flows to strong coupling, {\em i.e.},
leads to a triplet state between the impurity and the
conduction electrons.
At finite particle-hole asymmetry, there is still a transition between
a local moment and ferromagnetic strong coupling phase,
and the critical behavior is governed by a single critical
fixed point at finite asymmetry, ACR,
see Fig.~\ref{figflow}c.
The above changes do not influence the structure of the
flow diagram on the antiferromagnetic side.

\begin{figure}[t]
\epsfxsize=2.9in
\centerline{\epsffile{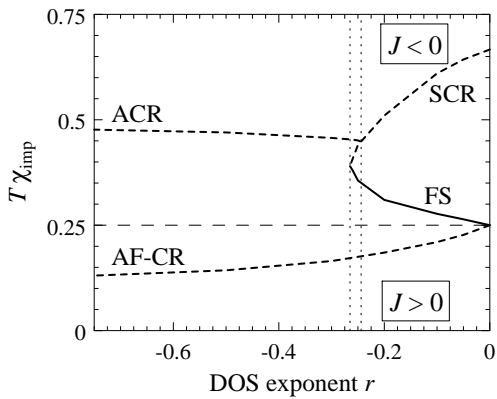}}
\vspace*{3pt}
\caption{
Values of $T\chi_{\rm imp}$ at the intermediate-coupling
fixed points deduced from
NRG \protect\cite{nrgprob} as function of the DOS exponent $r$.
Solid: Stable fixed point [fractional spin phase (FS)].
Dashed: Critical fixed points [ferromagnetic symmetric [SCR] and asymmetric [ACR]
as well as antiferromagnetic asymmetric (AF-CR)].
The horizontal dashed line separates the antiferromagnetic
($T\chi_{\rm imp}<0.25$) from the ferromagnetic regime ($T\chi_{\rm imp}>0.25$).
The vertical dotted lines mark the values of $\bar{r}^\ast \approx
-0.245$ and $\bar{r}_{\rm max} \approx -0.265$.
}
\label{figfp1}
\end{figure}

The Curie part of the impurity susceptibility, $T\chi_{\rm imp}$,
of the various intermediate-coupling fixed points is
shown in Fig.~\ref{figfp1}.
This illustrates
nicely the splitting of the unstable ferromagnetic fixed point
at $r = {\bar r}^\ast$ and the collapse of the
two fixed points at $r = {\bar r}_{\rm max}$.

The ferromagnetic side of the flow diagrams in
Fig.~\ref{figflow} shows some superficial similarity
to the antiferromagnetic $r>0$ situation \cite{GBI}:
Also in this case, two critical fixed points coexist over a
certain range of DOS exponents, $r^\ast < r < r_{\rm max}$.
However, there are crucial differences between the
present $r<0$, $J<0$ case and the $r>0$, $J>0$ regime
studied in Ref.~\onlinecite{GBI}:
First, the novel stable intermediate-coupling fractional-spin phase (FS)
has no counterpart for $r>0$.
Second, in the $r<0$ case the weak (strong) coupling fixed points are unstable (stable)
w.r.t. to particle-hole symmetry breaking; for $r>0$ this is reversed.

\begin{figure}[t]
\epsfxsize=2.9in
\centerline{\epsffile{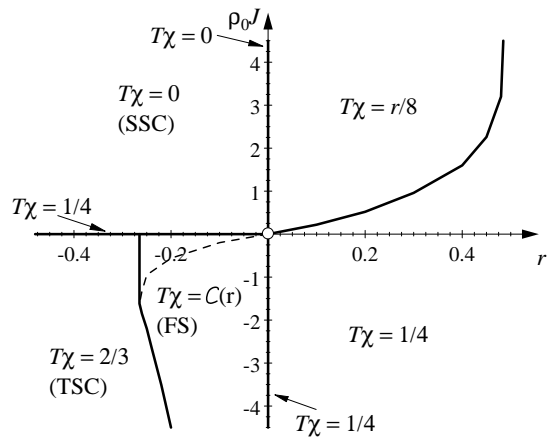}}
\vspace*{5pt}
\caption{
Phases of the symmetric power-law Kondo model in the
$(\rho_0 J)-r$ plane ($V\!=\!0$), with
the values of $T \chi_{\rm imp}$
characterizing each phase.
The $r>0$ sector has been explored, {\em e.g.}, in
Refs.~\protect\cite{withoff,ingersent,GBI};
the $r<0$ sector is covered by the present results.
Phase transitions (at fixed $r$) are tuned by moving along
vertical lines; all transitions at finite $J$ are continuous (i.e., of second order);
the $J=0$, $r=0$ point is the well-known Kosterlitz-Thouless
transition of the metallic Kondo model.
Solid lines are phase boundaries;
the dashed line within the fractional-spin phase (FS)
denotes the location of the intermediate-coupling fixed
point (estimated from NRG).
}
\label{figpd1}
\end{figure}

Fig.~\ref{figpd1} summarizes the phase diagram for the
particle-hole symmetric power-law Kondo model for both
signs of the exponent $r$ and both signs of $J$.
The line through the origin describes the fixed points
which can be deduced from the weak-coupling equations
(\ref{rgeq}) -- note that the line of critical fixed points
for $r>0$ has the same initial slope as the line of stable
fixed points for $r<0$.

In the vicinity of each critical point one can define an energy scale
$T^*$, which vanishes at the transition, and defines the crossover
energy above which quantum-critical behavior is observed~\cite{book}.
We note, however, that on the ferromagnetic side
simple one-parameter scaling
(as function of $T/T^*$ and $\omega/T^*$) can only be expected
for $V=0$ and ${\bar r}_{\rm max} < r < 0$ and
for $V\neq 0$ and $-1 < r < {\bar r}^\ast$.
In contrast, in the asymmetric case for ${\bar r}^\ast < r < 0$
the $J\!<\!0$ critical behavior is governed by the symmetric
multicritical point (SCR, Fig.~\ref{figflow}a), leading to
two distinct energy scales vanishing at the transition.
(This situation is similar to the behavior for small
positive $r$ \cite{GBI}.)
We have numerically checked the scaling behavior of
$\chi_{\rm imp}$ and of the T matrix,
details will appear elsewhere.


\section{Conclusions and outlook}

In this paper, we have examined a novel part of the phase diagram of the power-law Kondo model,
namely the behavior occuring for a bath density of states diverging as $|\omega|^r$ with exponent
$-1<r<0$. We have found rich physics for both antiferromagnetic and ferromagnetic Kondo coupling.
Particularly interesting is a stable phase with fractional ground state spin, {\em i.e.}, a Curie
susceptibility (\ref{curie}) with a non-trivial universal coefficient ${\cal C}(r)$, which also
shows a non-zero ground state entropy.
This phase occurs for ferromagnetic Kondo coupling and
corresponds to an intermediate-coupling fixed point in the renormalization
group sense.
It cannot be understood in a weak- or strong-coupling picture in terms of
the coupling between impurity and conduction electron degrees of
freedom.
Instead, it shares many properties of critical fixed points,
namely the absence of quasiparticle excitations,
scale invariance of the low-energy correlations,
and non-trivial power laws in response functions
like the dynamical susceptibility.
A preliminary analysis of the NRG levels
at the fractional-spin fixed point shows that
it {\em cannot} be described in terms of non-interacting
particles -- in contrast to, {\em e.g.}, the
two-channel Kondo fixed point which allows
for a Majorana fermion description \cite{bullafp}.

In closing, we mention that the discussed power-law Kondo
model is not only of purely theoretical interest:
Besides describing ``real'' magnetic impurities --
where a power law DOS can occur due to band structure
effects such as van-Hove singularities --
effective single-impurity models arise in the context of
lattice models in the limit of infinite dimensions
(dynamical mean-field theory (DMFT) \cite{DMFT}).
Here, the DOS of the effective embedding medium is determined
from a self-consistency condition involving both the
impurity spectrum and the bare DOS of the conduction band.
The interplay of these quantities can {\em e.g.} generate a
power-law DOS for the effective medium at a critical point
of the lattice model.
Particularly interesting in this context is the so-called
extended DMFT scheme \cite{EDMFT} where the influence of spin fluctuations
can drive the local impurity problem critical, which in turn
will generate a power-law spectral density of the effective fermionic
medium in extended DMFT.

Furthermore, we note that the low-energy physics of Kondo models is
related to that of dissipative two-level systems, i.e.,
a Kondo model with a diverging power-law DOS corresponds to
a spin-boson model with a certain non-ohmic bath.
Therefore, applications in mesoscopics, like two-level systems
coupled to a noisy environment, or in glass physics
may be found.

These prospects, together with the possibility of non-Fermi-liquid
physics in lattice realizations of the power-law Kondo physics,
e.g., in extended DMFT,
will be investigated in the future.


\acknowledgements

We thank M. Glossop, S. Kehrein, and Th. Pruschke
for valuable discussions.
This research was supported by the DFG through SFB 484.




\begin{references}


\bibitem{hewson} A.~C.~Hewson,
{\em The Kondo Problem to Heavy Fermions}, Cambridge
University Press, Cambridge (1997).

\bibitem{withoff} D.~Withoff and E.~Fradkin, Phys. Rev. Lett. {\bf
64}, 1835 (1990)

\bibitem{largeN} L.~S.~Borkowski and P.~J.~Hirschfeld, Phys.
Rev. B {\bf 46}, 9274 (1992);
G.-M.~Zhang {\em et al.}, \prl {\bf 86}, 704 (2001).

\bibitem{cassa} C.~R.~Cassanello and E.~Fradkin,
Phys. Rev. B {\bf 53}, 15079 (1996) and {\bf 56}, 11246 (1997).

\bibitem{chen} K.~Chen and C.~Jayaprakash, J. Phys.: Condens.
Matter {\bf 7}, L491 (1995).

\bibitem{bulla} R.~Bulla, T.~Pruschke,
and A.~C.~Hewson, J. Phys.: Condens. Matter {\bf 9}, 10463 (1997),
R.~Bulla, M.~T.~Glossop, D.~E.~Logan, and T.~Pruschke,
{\em ibid} {\bf 12}, 4899 (2000).

\bibitem{ingersent} K.~Ingersent, \prb {\bf 54}, 11936 (1996);
K.~Ingersent and Q.~Si, cond-mat/0109417.

\bibitem{GBI} C.~Gonzalez-Buxton and K.~Ingersent, \prb
{\bf 57}, 14254 (1998).


\bibitem{bulla2} M. Vojta and R.~Bulla, Phys. Rev. B {\bf 65}, 014511 (2002).

\bibitem{pgko} M. Vojta, \prl {\bf 87}, 097202 (2001).

\bibitem{impmag} S.~Sachdev, C.~Buragohain, and M.~Vojta,
Science {\bf 286}, 2479 (1999); M.~Vojta, C.~Buragohain, and
S.~Sachdev, Phys. Rev. B {\bf 61}, 15152 (2000).

\bibitem{NB} P. Nozi{\`e}res and A. Blandin,
J. Phys. (Paris) {\bf 41}, 193 (1980).

\bibitem{bullafp} R.~Bulla, A.~C.~Hewson, and G.-M.~Zhang,
\prb {\bf 56}, 11721 (1997), and references therein.

\bibitem{poor} P.~W.~Anderson, J. Phys. C {\bf 3}, 2436 (1970).

\bibitem{nrg} K.~G.~Wilson, Rev. Mod. Phys. {\bf 47}, 773 (1975).

\bibitem{nrgprob}
The values of $T \chi_{\rm imp}$ are sensitive to
the number of states, $N_s$, kept in each NRG step.
Most data shown are for $N_s=650$ and a discretization
parameter $\Lambda=2$.

\bibitem{book} S.~Sachdev, {\it Quantum Phase Transitions},
Cambridge University Press, Cambridge (1999).

\bibitem{DMFT} W.~Metzner and D.~Vollhardt, \prl {\bf 62}, 324 (1989);
A.~Georges, G.~Kotliar, W.~Krauth and M.~J.~Rozenberg,
Rev. Mod. Phys. {\bf 68}, 13 (1996).

\bibitem{EDMFT} Q. Si and L. Smith, \prl {\bf 77}, 3391 (1996);
Q. Si, S. Rabello, K. Ingersent, and L. Smith,
Nature {\bf 413}, 804 (2001), cond-mat/0202414;
for applications to the $t-J$ model see
O. Parcollet and A. Georges, \prb {\bf 59}, 5341 (1999);
K. Haule, A. Rosch, J. Kroha, and P. W\"olfle, cond-mat/0205347.

\end{references}
\end{document}